# Beyond Access: Guided LLM Scaffolding for Independent Learning in Undergraduate Statistics

**Mohammad Amanlou[a]*, Yasaman Amou-Jafari[a]†, Fereshte Bagheri[a]†, Fatemeh Boloukazari[a]†, Mehrad Liviyan[a]†, Elahe Khodaverdi Nadrabadi[a], Shahab Sherafat[a], Behnam Bahrak[b]**

[a]*School of Electrical and Computer Engineering, University of Tehran, Iran*

[b]*Tehran Institute for Advanced Studies, Khatam University, Iran*

***† These authors contributed equally and share second authorship.***

*mohammad.amanlou@ut.ac.ir

**Abstract:** Large language models (LLMs) are increasingly entering students' learning practices, but their educational value depends on whether they are used to support reasoning or to complete tasks without engaging in the underlying reasoning. This study examines guided LLM use in an undergraduate Probability and Statistics course, focusing on the distinction between assigned LLM access and the quality of students' actual interaction with the model. In a four-week quasi-experimental summer program, students were organized into three balanced conditions: no LLM access, unrestricted LLM access, and guided LLM access. The guided condition used the same LLM platform as the unrestricted condition, but students received explicit training and rules intended to promote reasoning-focused help-seeking, stepwise hints, verification, and ethical use. All quizzes and the delayed final exam were completed without LLM or external assistance, allowing us to separate AI-supported practice performance from independent learning. Results show that guided use was associated with a clearer learning-oriented interaction pattern than unrestricted access, especially in prioritizing reasoning over final answers and requesting stepwise support. In behavior-defined analyses, Guided-LLM students showed a promising pattern of stronger no-help quiz performance, while practice scores showed no consistent Guided-LLM advantage. Available time measures did not support a simple duration-based explanation, and self-assessment calibration suggested better alignment between perceived and demonstrated understanding in Guided-LLM. In this setting, access alone did not reliably distinguish independent performance; the observed process and outcome patterns instead motivate further testing of scaffolds that encourage reasoning-focused LLM use. For AIED, these findings provide preliminary evidence for evaluating LLM-supported learning in terms of interaction quality as well as access.

## 1. Introduction

Large language models (LLMs) are rapidly changing how students engage with academic work, from clarifying concepts and generating examples to checking solutions and receiving assistance. Evidence suggests that ChatGPT-supported learning can improve outcomes on average, but effects vary across instructional contexts, implementation designs, and patterns of student use (Mo et al., 2026). Reviews of generative AI in education similarly emphasize a dual possibility: LLMs can increase engagement, support, and access to explanations, but may also encourage overreliance, reduced critical thinking, academic misconduct, or superficial task completion when used without appropriate pedagogical structure (Kasneci et al., 2023; Lo et al., 2024).

This tension makes LLM integration an instructional-design problem rather than a simple question of access. The same model can function as an answer-seeking tool or as a tutor-like partner that supports reasoning, self-explanation, and metacognitive monitoring. Recent study-oriented LLM interfaces reflect this shift toward hints and step-by-step guidance (OpenAI, 2025), but controlled classroom evidence remains limited on whether explicit guidance changes students' LLM-use behavior and whether such changes translate into stronger independent performance when LLM assistance is removed.

We examine this question in a foundational Probability and Statistics course, where topics such as conditional probability, Bayesian reasoning, independence, and distributions require careful interpretation of assumptions, symbolic reasoning, and resistance to intuitive but incorrect judgments (Díaz & Batanero, 2009). In a four-week quasi-experimental summer program with second-semester Electrical Engineering and Computer Engineering undergraduates at University of Tehran, all students received the same instructional videos, practice assignments, no-help quizzes, and delayed final exam. After a Week 1 run-in phase, students were organized into three balanced conditions: No-LLM, Unrestricted-LLM, and Guided-LLM; from Topics 2–4, the groups differed only in practice-time LLM policy, while all assessments remained no-help.

The study asks whether guided LLM use during practice is associated with stronger independent performance and with differences in time-on-task, chat behavior, rule-following, and self-assessment calibration. Its contribution is a quasi-experimental three-condition classroom study that separates AI-supported practice from no-help assessment and links outcome measures with transcript-based and self-report process evidence. We first examine whether guidance was associated with enacted-use differences, then whether these behavior-defined differences corresponded to no-help performance, and finally whether time-on-task or calibration offered alternative explanations.

## 2. Related Work

Large Language Models (LLMs) have become common in student learning, but evidence about their educational value remains mixed. Early work argued that education should move beyond prohibition toward responsible integration while managing risks such as hallucination, bias, and brittle reasoning (Kasneci et al., 2023). Recent reviews likewise suggest that LLM-supported learning is neither uniformly beneficial nor harmful; its effects depend on instructional design, assessment context, and whether students use LLMs for reflection or answer-seeking (Kasneci et al., 2023; Lo et al., 2024). Although LLM-supported instruction can improve immediate performance in domains such as medical education and language learning (Zeng et al., 2025; Z. Zhang & Huang, 2024), such gains do not necessarily imply durable conceptual learning or retention (Kalam et al., 2025; Zhu et al., 2025). This makes it necessary to distinguish immediate AI-supported performance from durable, independently demonstrated understanding.

A central risk is that unguided LLM access may produce superficial success: students complete tasks efficiently while developing weaker independent capability. Prior work suggests that unrestricted access can reduce engagement (Nie et al., 2025), impair retention when used as a cognitive crutch (Barcaui, 2025), and harm learning when students seek direct solutions rather than explanations (Lehmann et al., 2024). This risk aligns with learning-science accounts of effortful retrieval, self-explanation, and productive struggle (Bjork & Bjork, 2020; Chi et al., 1994; Kapur, 2008; Roediger & Karpicke, 2006), as well as Intelligent Tutoring Systems research on hint abuse, gaming, and maladaptive help-seeking (Aleven et al., 2016; Baker et al., 2004). Because LLMs can act as tutor, hint system, answer generator, or feedback source, their educational role is enacted through interaction. Guidance is therefore theoretically grounded in scaffolding, the zone of proximal development, self-regulated learning, and digital prompting research, all of which emphasize preserving learner agency while supporting progress toward independent performance (Thomann & Deutscher, 2025; Vygotsky, 1978; Wood et al., 1976; Zimmerman, 2002).

LLM-specific evidence reinforces the need for such scaffolding: LLMs can support learning workflows such as generating difficulty-controlled questions (Amanlou et al., 2026), but students do not automatically use them. Students often default to direct-instruction prompts rather than reflective or higher-order inquiry (Mutanga et al., 2025), whereas guided LLM use can improve engagement and writing quality relative to unguided access (M. Zhang, 2025), consistent with help-seeking research showing that learners benefit from explicit instruction on requesting assistance (Aleven et al., 2016; Roll et al., 2004). Guidance is also needed because fluent but incorrect LLM outputs can introduce misinformation (Rahman et al., 2026). Educational LLM use therefore requires AI literacy and calibrated trust: learners must know when to rely on, question, and verify model outputs (Hoff & Bashir, 2015; Long & Magerko, 2020). UNESCO guidance similarly emphasizes human-centered design, critical engagement, and verification rather than passive acceptance of generated responses (Miao & Holmes, 2023), making verification part of the learning mechanism rather than an integrity safeguard.

Methodologically, evaluating LLM-supported learning requires process evidence in addition to outcome scores. Learning analytics shows that digital traces can capture multidimensional engagement (Tempelaar et al., 2020), while time-on-task measures require careful interpretation because they are sensitive to idle time and session boundaries (Kovanović et al., 2015). In LLM contexts, this is especially important because identical endpoint scores may reflect different interaction patterns: explanation-seeking and solution-seeking can have opposite learning implications (Lehmann et al., 2024), access can affect engagement in complex ways, and prompt types can indicate different levels of cognitive engagement (Mutanga et al., 2025). Taken together, this literature leaves a specific gap: controlled evidence linking guidance, enacted interaction quality, and no-help performance. The present study addresses this gap by combining no-help assessments with transcript-based rule-following, time measures, and self-assessment calibration. Guided prompting itself is not our claimed novelty. We operationalize established scaffolding, help-seeking, self-regulation, active-practice, and AI-literacy principles as six observable LLM-use rules, verify their enactment from transcripts, and relate behavior-defined profiles to subsequent no-help performance.

## 3. Study Context and Experimental Design

### 3.1 Course Context, Participants, and Ethics

The study was conducted in a controlled four-week summer program on Probability and Statistics for second-semester Electrical Engineering and Computer Engineering undergraduates at University of Tehran. The course covered foundational probabilistic reasoning, including conditional probability, Bayes' theorem, random variables, distributions, and joint distributions, which can remain conceptually difficult after formal instruction (Díaz & Batanero, 2009). Participation was voluntary, did not affect official grades, and required written informed consent covering the research use of LLM chats, scores, timestamps, and activity records. Raw records were retained on the course server under the study consent, and only de-identified, group-level results are reported. The study was conducted with the required course-level permissions. Instruction was delivered asynchronously in the students' native language through official videos prepared by the course instructor. The summer format reduced competing academic demands and enabled a repeated weekly cycle of instruction, practice, and no-help assessment.

### 3.2 Grouping Procedure and Conditions

The study used a shared Week 1 run-in phase followed by a three-condition intervention from Topic 2 onward. Of the 57 enrolled students, seven did not complete the Week 1 run-in and were therefore not assigned to an intervention condition. The remaining 50 students had Week 1 baseline scores and were allocated through stratified manual balancing based on a composite of Practice 1 and Quiz 1, so that higher-, middle-, and lower-performing students were represented across the three conditions (No-LLM: 17; Unrestricted-LLM: 17; Guided-LLM: 16). From Topic 2 onward, groups differed in practice-time LLM access and guidance:

No-LLM completed practice without LLMs; Unrestricted-LLM could freely use the course LLM during practice, subject to standard academic-integrity expectations; and Guided-LLM used the same LLM through the same platform after explicit training and rules for learning-oriented use. All quizzes and the delayed final exam were administered under a strict no-help policy: no LLMs, external resources, or peer assistance were permitted. This quasi-experimental design separated AI-supported practice from no-help assessment, allowing us to distinguish assisted task completion from independently demonstrated learning.

### 3.3 Learning Tasks, Assessment, and Scoring

Each weekly unit followed the same structure: students watched the assigned lecture videos, completed a six-question open-ended practice assignment, with some questions containing multiple subparts, and then took a no-help quiz. Each quiz included five multiple-choice items and three open-response questions, some with subparts. The course concluded with a cumulative 11-question no-help final exam two weeks after Quiz 4 to assess delayed retention.

Open responses were blind-graded: student identities and group labels were hidden, and each item was graded consistently across students. Five graders contributed to the scoring workflow. All final submitted answers were handwritten on paper, reducing direct copy-paste from LLM output and requiring students to reconstruct their solutions manually even when LLM use was allowed during practice.

### 3.4 LLM Platform and Guided-Use Protocol

All LLM activity for the two LLM-access groups occurred through a centralized course platform using the gpt-4o-mini model through an OpenAI-compatible API. The same model, interface, and platform capabilities were used in both LLM-access conditions; Guided-LLM differed through student-facing training and rules rather than a separate model or hard-coded guided mode. The platform provided no browsing, plugins, or alternative models, logged prompts, responses, timestamps, user identifiers, and interaction histories, and used prior interactions to maintain continuity across sessions. The platform and reproducibility materials are publicly available.[1]

Before Topic 2, Guided-LLM students completed a slide-based orientation with examples of productive and unproductive prompts, then received weekly reminders during Topics 2–4. The training framed the model as a fallible tutor rather than an answer source and emphasized explanation-seeking, stepwise help, verification, and ethical use. This emphasis aligns with responsible generative-AI guidance that stresses human-centered design, critical engagement, and verification rather than passive acceptance of model outputs (Miao & Holmes, 2023).

The protocol defined six observable rule families: R1 process-over-answer and R3 stepwise-hint seeking (productive help-seeking/scaffolding), R2 concept tutoring, R4 active practice and retrieval, R5 verification and calibrated trust, and R6 ethical/no-help use (Aleven et al., 2016; Bjork & Bjork, 2020; Long & Magerko, 2020; Zimmerman, 2002). This operationalization translates established learning-science principles into transcript-codeable LLM-use behaviors rather than proposing a generic prompting template.

### 3.5 Data Sources, Compliance Coding, and Analysis Strategy

The study combined outcome, process, chat, and self-report data. Performance outcomes included practice, quiz, and final-exam scores, all normalized to a 0–100 scale. Practice time was self-reported, whereas quiz and final-exam times were automatically logged from question start to submission. Students rated their understanding after each topic on a 1–10 scale, with an additional final rating for overall course understanding. For calibration analyses, the 1–10 self-assessment ratings were linearly rescaled to 0–100 by multiplying by 10 before subtracting matched no-help performance.

[1] GitHub repository: https://github.com/MohammadAmanlou/Guided-LLM-Scaffolding

We directly measured guided use instead of assuming it from group assignment alone. Three raters independently coded 67 available student-practice transcripts from Practices 2–4 across 32 LLM-access students. For each record, raters judged whether each of the six rule families was satisfied. Inter-rater agreement before consensus was high (overall Fleiss' $\kappa$ = .896; observed agreement = 92.0%), with rule-specific $\kappa$ values ranging from .871 to 0.99. Disagreements were resolved by majority vote, using the label supported by at least two raters, producing a rule-following score from 0 to 6.

Because the study focused on students' enacted LLM-use behavior, outcome analyses used a behavior-defined grouping derived from transcript-based compliance scores. Two students originally assigned to Unrestricted-LLM were classified as guided-like, and two students originally assigned to Guided-LLM were classified as unguided-like. This reclassification was determined solely from chat behavior and was finalized before practice, quiz, and final-exam outcome statistics were examined. Because the enacted grouping was nevertheless defined using post-intervention behavior, comparisons based on these groups are interpreted as associational rather than as causal estimates of assignment effects.

For statistical reporting, we emphasize descriptive patterns, effect sizes, and robustness checks rather than treating single p-values as definitive evidence. For three-group outcome comparisons, we report group means and standard deviations together with one-way ANOVA p-values and Kruskal–Wallis p-values as non-parametric robustness checks. For two-group compliance comparisons between the LLM-access conditions, we report student-level mean differences and two-sided test p-values where applicable. Rule-level compliance rates were compared as binary proportions. Because multiple outcomes and process indicators were examined, p-values were not treated as confirmatory evidence. For key pairwise outcome contrasts, we report mean differences, Hedges' g, Welch p-values, and percentile-bootstrap 95% confidence intervals based on 2,000 within-group resamples. Pairwise results were interpreted as exploratory.

**Data and materials availability.** Platform code, study configuration, guided-use materials, prompt examples, coding guide, analysis scripts, and de-identified interaction excerpts are publicly available in the repository. Full transcripts and participant-level assessment records are withheld because they contain learner data covered by consent and privacy constraints.

## 4. Results

### *4.1 Participant Flow and Initial Comparability*

The study began with 57 students. Seven did not complete the Week 1 run-in and were therefore not assigned to an intervention condition. The remaining 50 students had Week 1 scores and were allocated to No-LLM ($n = 17$), Unrestricted-LLM ($n = 17$), or Guided-LLM ($n = 16$). Forty-nine contributed at least one subsequent practice or quiz record, and 37 met the final analytic inclusion criterion after excluding students who missed two or more topic cycles (No-LLM: 13; Unrestricted-LLM: 12; Guided-LLM: 12). Post-assignment attrition was therefore 4/17, 5/17, and 4/16, respectively, with no evidence that attrition differed across conditions (Fisher–Freeman–Halton exact $p = 1.000$). The final analytic sample remained broadly comparable on Week 1 Practice and Quiz scores. Table 1 reports this balance for the final outcome-analysis sample. Neither Practice 1 nor Quiz 1 showed a statistically significant group difference, and the largest pairwise standardized mean difference after filtering was 0.47, suggesting no large observed pre-intervention imbalance.

Table 1. *Week 1 comparability in the final analytic sample.* Scores are on a 0–100 scale; values are Mean (SD). LLM = large language model; SD = standard deviation; ANOVA = analysis of variance; F = ANOVA test statistic; η² = eta-squared effect size.

| Week 1 measure | No-LLM Mean (SD) | Unrestricted-LLM Mean (SD) | Guided-LLM Mean (SD) | Three-group test |
|---|---|---|---|---|
| Practice 1 | 58.11 (18.78) | 59.50 (23.47) | 54.54 (21.14) | $F = 0.18$, $p = .839$, $\eta^2 = .010$ |
| Quiz 1 | 66.92 (23.58) | 56.66 (19.46) | 60.00 (23.54) | $F = 0.69$, $p = .508$, $\eta^2 = .039$ |

## 4.2 Guidance and Enacted LLM Use

We next examined whether assigned guidance was associated with different enacted LLM use. This manipulation check compared the two LLM-access groups across Practices 2–4, coding each student-practice record against six rule families.

Guided-LLM students showed higher overall compliance than Unrestricted-LLM students under assigned groups. The behavior-based enacted grouping, finalized from transcript compliance before outcome statistics were examined, was subsequently used to characterize observed LLM-use profiles and compare learning outcomes. Rule-level results showed that the contrast was driven mainly by process-oriented help-seeking and stepwise hints, whereas conceptual explanation was common in both groups. Active-learning and verification behaviors were only partially adopted, suggesting that higher-agency strategies may require stronger scaffolding. Exploratory chat-process indicators were consistent with this interpretation: Guided-LLM sessions contained more user turns on average (17.21 vs. 10.54) and more user text (2,645 vs. 1,410 characters). Because the primary manipulation check is transcript-coded rule following, these chat-process indicators are reported descriptively. Overall, the transcript-coded patterns indicate qualitatively different forms of LLM use, not merely different levels of LLM access. Because the enacted grouping was itself defined from post-intervention compliance behavior, the larger enacted-group compliance separation is partly mechanical and is not interpreted as an independent treatment effect.

Table 2. *Rule-following compliance among LLM-access students.* values are mean satisfied rule families out of six; rule-level values are compliance percentages. Assigned groups use original intervention labels; final/enacted rows use behavior-based labels and are exploratory. pp = percentage points; d = Cohen's d; p-values are two-sided.

| Indicator | Unrestricted | Guided | Result |
|---|---|---|---|
| Overall compliance, assigned | 2.43 / 6 | 3.21 / 6 | **+0.78 / 6; d = 0.81; p = .025** |
| Overall compliance, enacted | 2.18 / 6 | 3.73 / 6 | **+1.55 / 6; exploratory sensitivity** |
| R1: Process over answers | 22% | 100% | **+78 pp; p < .001** |
| R2: Concept tutoring | 96% | 100% | +4 pp; p = 1.000 |
| R3: Stepwise hints | 7% | 41% | **+33 pp; p = .009** |
| R4: Active learning | 11% | 26% | +15 pp; p = .293 |
| R5: Critical evaluation | 30% | 41% | +11 pp; p = .569 |
| R6: Ethical/no-help use | 59% | 70% | +11 pp; p = .569 |
| Practice 2 compliance | 1.82 / 6 | 3.73 / 6 | **+1.91 / 6; p < .001** |
| Practice 3 compliance | 2.62 / 6 | 3.89 / 6 | **+1.26 / 6; p = .007** |
| Practice 4 compliance | 2.50 / 6 | 3.71 / 6 | **+1.21 / 6; p = .015** |

## 4.3 Enacted LLM Use and Independent Performance

Having characterized differences in enacted LLM use, we next examined whether these behavior-defined group differences corresponded to learning outcomes during assisted practice and independent no-help assessment. Practice performance showed no consistent Guided-LLM advantage: none of the Practice group comparisons reached statistical significance, and the highest Practice 4 mean appeared in the Unrestricted-LLM group. This suggests that assisted practice scores should not be treated as direct evidence of retained independent learning.

In contrast, no-help quiz performance showed a promising but exploratory pattern. Guided-LLM had the highest mean in all three intervention quizzes, with the strongest separation in Quiz 4. Exploratory pairwise contrasts favored Guided-LLM over No-LLM by 23.17 points (95% bootstrap CI [5.29, 39.49], Hedges' g = 0.95, Welch p = .022) and over Unrestricted-LLM by 16.42 points (95% bootstrap CI [−0.42, 32.84], Hedges' g = 0.73, Welch p = .079). The Quiz 4 omnibus evidence was significant under the Kruskal–Wallis test but marginal under ANOVA. The final exam followed the same descriptive direction, but uncertainty remained substantial: Guided-LLM exceeded No-LLM by 14.37 points (95% bootstrap CI [−2.69, 31.07], Hedges' g = 0.61, Welch p = .127) and Unrestricted-LLM by 14.24 points (95% bootstrap CI [−0.99, 28.67], Hedges' g = 0.71, Welch p = .086). Because both final-exam intervals included zero and the omnibus tests were not significant, these results should be interpreted as directional rather than statistically confirmatory.

Table 3. *Practice, no-help quiz, and final-exam outcomes for the behavior-defined enacted groups.* Values are Mean (SD); ANOVA p and KW p report omnibus three-group tests, with KW denoting Kruskal–Wallis. Key pairwise mean differences, Hedges' g values, Welch p-values, and 95% bootstrap confidence intervals are reported in the text.

| Outcome | No-LLM (SD) | Unrestricted-LLM (SD) | Guided-LLM (SD) | ANOVA p | KW p |
|---|---|---|---|---|---|
| Practice 2 | 52.62 (18.05) | 53.42 (16.74) | 59.00 (17.43) | .619 | .543 |
| Quiz 2 | 65.45 (26.57) | 57.10 (22.59) | 67.04 (15.24) | .501 | .389 |
| Practice 3 | 49.69 (16.42) | 51.46 (12.32) | 53.00 (16.44) | .863 | .875 |
| Quiz 3 | 53.38 (30.30) | 57.83 (31.54) | 61.67 (19.46) | .758 | .659 |
| Practice 4 | 51.54 (23.41) | 66.00 (26.23) | 51.75 (25.18) | .274 | .260 |
| Quiz 4 | 50.08 (26.47) | 56.83 (23.31) | 73.25 (20.20) | .055 | .045 |
| Final exam | 56.03 (25.23) | 56.16 (18.88) | 70.40 (19.94) | .185 | .177 |

## *4.4 Time and Calibration*

Finally, we examined whether the outcome pattern could be explained by time-on-task or self-assessment differences rather than by the quality of LLM use. Time-on-task did not support a simple duration-based explanation for the outcome pattern. Although Unrestricted-LLM reported the lowest average practice time, no topic-level time comparison was statistically significant. Guided-LLM also reported less practice time than No-LLM on average, while achieving stronger no-help performance. Logged quiz and final-exam times were similar across groups. Thus, the available time measures do not support a simple duration-based explanation, although self-reported practice time should be interpreted cautiously.

Self-assessment calibration provided an additional process-level check. Calibration gaps were computed as self-assessment minus matched no-help performance; positive values indicate overconfidence and negative values indicate conservative self-assessment. Students were classified as well-calibrated when this gap fell within a ±10-point practical tolerance band. Guided-LLM showed the strongest alignment between perceived and demonstrated understanding, with the highest well-calibrated rate and the strongest self-assessment–performance correlations. In contrast, Unrestricted-LLM tended toward conservative self-assessment, whereas No-LLM showed the clearest overestimation pattern.

Table 4. *Time-on-task and self-assessment calibration* (Time values are in minutes. Calibration gap = self-assessment minus no-help performance; positive values indicate overconfidence. Well-calibrated means within ±10 points. KW = Kruskal–Wallis; r = Pearson correlation).

| Indicator | No-LLM | Unrestricted-LLM | Guided-LLM | Overall evidence |
|---|---|---|---|---|
| Practice time mean | 101.43 | 71.65 | 87.19 | No significant difference |
| Quiz time mean | 86.78 | 88.37 | 86.20 | No significant difference |
| Final-exam time | 90.10 | 94.27 | 94.34 | No significant difference |
| Topic 2 calibration gap | −6.95 | −15.21 | −7.32 | ANOVA p = .342; KW p = .440 |
| Topic 3 calibration gap | +2.20 | −4.92 | −6.64 | ANOVA p = .582; KW p = .339 |
| Topic 4 calibration gap | +17.42 | −14.36 | −1.00 | **ANOVA p < .001; KW p < .001** |
| Final calibration gap | +7.74 | −8.32 | +3.96 | **ANOVA p = .055; KW p = .038** |
| Overconfident rate | 33% | 19% | 21% | Lowest in Unrestricted-LLM |
| Well-calibrated rate | 33% | 39% | 43% | Highest in Guided-LLM |
| Underconfident rate | 33% | 42% | 36% | Highest in Unrestricted-LLM |
| Topic self-assessment vs. matched quiz, r | .68 | .52 | .75 | **All: r = .62, p < .001** |
| Final self-assessment vs. final exam, r | .75 | .25 | .90 | **All: r = .66, p < .001** |

## 5. Discussion

### *5.1 From LLM Access to Effective LLM Use*

The clearest process result is that assigned LLM access and enacted use were not equivalent. Although Unrestricted-LLM and Guided-LLM used the same platform, Guided-LLM showed more reasoning-focused help-seeking and stepwise hint use, whereas concept tutoring was common in both groups. For AIED, this supports measuring learner–LLM interaction quality alongside access; in this study, prompt-literacy guidance was used as an instructional scaffold rather than merely as technical instruction.

### *5.2 Assisted Practice Is Not the Same as Independent Learning*

Practice scores showed no consistent Guided-LLM advantage, whereas no-help quizzes showed the clearest descriptive separation in Quiz 4. This does not show that unrestricted use harmed learning; rather, it illustrates why AI-assisted practice performance should not be equated with independent learning. The final exam remained directional and non-confirmatory, so cumulative or task-sensitive interpretations require larger studies.

### *5.3 Calibration and Metacognitive Value*

Calibration results were exploratory. Guided-LLM showed the highest well-calibrated rate and strongest self-assessment–performance correlations, while Unrestricted-LLM tended toward underconfidence and No-LLM toward overconfidence. These patterns are consistent with, but do not establish, improved metacognitive monitoring or calibrated trust.

### *5.4 Implications for AIED and Course Design*

For AIED, the results motivate evaluating LLM-supported learning beyond binary access by measuring enacted interaction and preserving no-help assessments. Candidate scaffolds for future evaluation include prompt starters, hint-first workflows, reflection checkpoints, and verification cues. Because AI-assisted practice scores may reflect real-time support rather than retained competence, independent quizzes, written explanations, oral checks, or transfer tasks can provide complementary evidence of learning. These are design hypotheses from a small single-course study rather than general prescriptions.

### *5.5 Limitations and Future Work*

This study was conducted in a single summer course with a relatively small final analytic sample and substantial participation filtering, limiting statistical power and generalizability. Because assignment was not randomized and outcome analyses used behavior-defined post-intervention groups, findings are associational rather than causal, and enacted-group separation may be amplified by the grouping procedure. Process measures also had limitations: practice time was self-reported, transcript coding required human judgment despite multiple raters, and Guided-LLM received additional training and reminders that may have introduced attention, expectancy, or demand effects beyond the guidance content itself. Future work should test guided LLM use across domains, learner populations, models, and interface designs, using richer platform logs, larger randomized designs, and matched-attention controls.

## 6. Conclusion

This single-course study provides preliminary evidence that enacted LLM-use quality may matter alongside access. Guided-LLM students showed more reasoning-focused help-seeking and stepwise hint use, while behavior-defined analyses showed a promising later no-help quiz pattern. The delayed final exam followed the same descriptive direction but was not statistically confirmatory. Larger randomized studies are needed to test whether scaffolding LLM use toward reasoning, verification, and independent problem solving improves durable, independently demonstrated learning.